\documentclass[aps,prl,twocolumn,superscriptaddress,amsmath,amssymb]{revtex4}
\usepackage{graphicx}
\usepackage{amsmath}
\usepackage{pstricks}
\usepackage{tikz}
\usepackage{float}

\begin{document}

\title{Efimov physics and universal trimer in spin-orbit coupled ultracold atomic mixtures}
\author{Zhe-Yu Shi}
\affiliation{Institute for Advanced Study, Tsinghua University, Beijing, 100084, China}
\author{Hui Zhai}
\affiliation{Institute for Advanced Study, Tsinghua University, Beijing, 100084, China}
\author{Xiaoling Cui}
\email{xlcui@iphy.ac.cn}
\affiliation{Beijing National Laboratory for Condensed Matter Physics, Institute of Physics, Chinese Academy of Sciences, Beijing, 100190, China}
\date{\today}
\begin{abstract}
We study the two-body and three-body bound states in ultracold atomic mixtures with one of the atoms subjected to an isotropic spin-orbit (SO) coupling. 
We consider a system of two identical fermions interacting with one SO coupled atom. It is found that there can exist two types of three-body bound states, Efimov trimers and universal trimers. The Efimov trimers are energetically less favored by the SO coupling, which will finally merge into the atom-dimer threshold as increasing the SO coupling strength. Nevertheless, these trimers exhibit a new kind of discrete scaling law incorporating the SO coupling effect. On the other hand, the universal trimers are more favored by the SO coupling. They can be induced at negative s-wave scattering lengths and with smaller mass ratios than those without SO coupling. These results are obtained by both the Born-Oppenheimer approximation and exact solutions from three-body equations.
\end{abstract}
\maketitle

\section{\uppercase\expandafter{\romannumeral1}. introduction}

Few-body problems constitute an important part in the field of ultracold atoms. The study of two-body problem is fundamentally crucial for engineering strong interaction\cite{Feshbach}; while the study of three-body problem is practically useful for controlling the lifetime of atomic gases due to three-body recombination\cite{atom_loss}. In particular, the three-body system has very intriguing bound state properties that are closely related to atom loss. A typical example is the Efimov trimer, characterized by a sequence of three-body bound states whose binding energies obey a discrete scaling law at two-body scattering resonances\cite{Efimov,Braaten}. The Efimov physics has been successfully explored in cold atom experiments by detecting the enhanced three-body loss rate\cite{Efimov_Exp1,Efimov_Exp2,Efimov_Exp3,Efimov_Exp4,Efimov_Exp5,Efimov_Exp6,Efimov_Exp7,Efimov_Exp8,Efimov_Exp9,Efimov_Exp10,Efimov_Exp11}, the Efimov spectrum from radio-frequency spectroscopy\cite{Efimov_rf1,Efimov_rf2}, and the discrete scaling from the successive atom loss peaks\cite{Efimov_scaling1,Efimov_scaling2,Efimov_scaling3}. Moreover, another type of trimers exists in the mass-imbalanced two-component fermions. In contrast to the Efimov trimers, the binding energies of universal trimers universally depend on the s-wave scattering length regardless of short-range interaction details\cite{KM,Ueda}. These universal trimers have not been observed in cold atoms experiment yet, due to their special requirement on the mass ratio of different fermions.

Recently, the spin-orbit (SO) coupling has been successfully realized and studied in cold atom experiments using two-photon Raman processes
\cite{Spielman_exp1,Spielman_exp2,Shuai,Spielman_exp3,Jing,MIT,Chuanwei,Spielman_exp4,Shuai_2013,Spielman_2013,Jing_2013}. Though what is realized now is the equal mixing of Rashba and Dresselhaus SO coupling, it is conceivable to get access to other types of SO coupling with higher symmetry, such as the Rashba and isotropic types, given quite a number of proposals have been made along this direction\cite{rashba_proposal,spielman_3d_1,spielman_3d_2,Xu}. While extensive studies have been focused on the SO coupling effect to the two-body and many-body cold atom systems(see reviews \cite{review}), there have been very few studies on the three-body problem with SO coupling\cite{SCZ,prx}. These studies reveal the significant effects of SO coupling to the three-body bound states, in facilitating the formation of universal trimer on top of an atom-dimer background\cite{SCZ} and even inducing the universal Borromean binding without the formation of any two-body bound state\cite{prx}.

In this work, we extend our previous work\cite{SCZ} to investigate both the two-body and three-body bound states in ultracold atomic mixtures, when one of the atoms is subjected to an isotropic SO coupling. In this work we focus more on the properties of Efimov trimer in a SO coupled system. We give a qualitative understanding for these properties based on the Born-Oppenheimer approximation and we present more details about the derivation of two-body and three-body equations from the Lippmann-Schwinger equation, as well as more comprehensive discussions on the effect of SO coupling to few-body bound states.

The organization of this paper is as follows. In Section \uppercase\expandafter{\romannumeral2}, we study the two-body bound state. In Section \uppercase\expandafter{\romannumeral3}, we use the Born-Oppenheimer approximation to get an intuitive picture of the SO coupling effects on three-body bound states. In Section \uppercase\expandafter{\romannumeral4}, we exactly solve the three-body bound states with the zero total momentum. We show the results of Efimov trimer in subsection A, and universal trimer in subsection B. Finally we summarize our results in Section \uppercase\expandafter{\romannumeral5}.

\section{\uppercase\expandafter{\romannumeral2}. Two-body system}
In this section, we consider a spinless $\alpha$-atom with mass $M$ interacting with a spin-$\frac{1}{2}$ $\beta$-atom with mass $m$ via a contact s-wave interaction. The two-body Hamiltonian reads (we set $\hbar=1$):
\begin{eqnarray}
H_{\text{2b}}=\frac{\mathbf{p_1}^2}{2M}+\frac{\mathbf{p_2}^2}{2m}-\frac{\lambda\mathbf{p_2}\cdot\hat{\sigma}}{m}
+g\delta(\mathbf{r_1-r_2}).\label{2-body H}
\end{eqnarray}
Here $\hat{\sigma}$ is the spin operator of $\beta$-atom, which couples to its momentum via a three-dimensional isotropic SO coupling $\lambda \mathbf{p}\cdot\hat{\sigma}$. Without loss of generality we always assume the SO coupling strength $\lambda\geq0$. The coupling constant $g$ is invariant under spin rotation and related to the $s$-wave scattering length $a$ by
\begin{eqnarray}
\frac1g=\frac{Mm}{2\pi(M+m)a}-\frac1\Omega\sum_{\mathbf{k}}\frac{2Mm}{(M+m)k^2},
\end{eqnarray}
where $\Omega$ is the system volume.

Before calculating the bound state energy, we should define the two-atom threshold energy at first. Since the total momentum $\mathbf{K}$ is always a good quantum number of the two-body Hamiltonian, we can define the two-atom threshold energy as the ground state energy of two non-interacting particles with total momentum $\mathbf{K}$,
\begin{eqnarray}
E_\text{aa}(\mathbf{K})&=&\min_{\mathbf{k_1+k_2=K},\pm}\bigg{(}\varepsilon_{\mathbf{k_1}}+\epsilon_{\mathbf{k_2}}^\pm\bigg{)}\nonumber\\
&=&\frac{(K-\lambda)^2}{2(M+m)}-\frac{\lambda^2}{2m}.
\end{eqnarray}
Here $\varepsilon_\mathbf{k}=k^2/2M$ is the dispersion for $\alpha-$atom. And $\epsilon^\pm_\mathbf{k}=(k^2\pm2|k|\lambda)/2m$ is the dispersion for SO coupled $\beta$-atom, $\pm$ represents two different helicity branches of the SO coupled $\beta$-atom. In the subspace with total momentum $\mathbf{K}$, the two-body bound state should be an eigenstate with energy lower than the two-atom threshold $E_{aa}(\mathbf{K})$.

We calculate the bound state energy through the Lippmann-Schwinger equation in momentum space,
\begin{eqnarray}
\Psi_\sigma(\mathbf{K-p,p})=\frac{g}{\Omega}\sum_{\mathbf{q},\sigma'}G^{(0)}_{\sigma,\sigma'}(\mathbf{K-p,p})\Psi_{\sigma'}(\mathbf{K-q,q}),\label{2-body eqn}
\end{eqnarray}
where $\Psi_{\sigma}(\mathbf{K-p,p})$ is the wavefunction in momentum space, $\sigma=\uparrow,\downarrow$ represents different spin components of $\beta$-atom, $\mathbf{p}$ is the momentum of $\beta$-atom and $\mathbf{K}$ is the total momentum. $G^{(0)}_{\sigma,\sigma'}$ is the Green's function for two noninteracting particles. It can be calculated that
\begin{eqnarray}
G_{\uparrow\uparrow}(\mathbf{k_1,k_2})&=&\frac{\cos^2\frac{\theta_\mathbf{k_2}}{2}}
{E-\varepsilon_{\mathbf{k_1}}-\epsilon^+_{\mathbf{k_2}}}+
\frac{\sin^2\frac{\theta_\mathbf{k_2}}{2}}
{E-\varepsilon_{\mathbf{k_1}}-\epsilon^-_{\mathbf{k_2}}},\\ \nonumber
G_{\downarrow\downarrow}(\mathbf{k_1,k_2})&=&\frac{\sin^2\frac{\theta_\mathbf{k_2}}{2}}
{E-\varepsilon_{\mathbf{k_1}}-\epsilon^+_{\mathbf{k_2}}}+
\frac{\cos^2\frac{\theta_\mathbf{k_2}}{2}}
{E-\varepsilon_{\mathbf{k_1}}-\epsilon^-_{\mathbf{k_2}}},\\ \nonumber
G_{\downarrow\uparrow}=G^*_{\uparrow\downarrow}&=&\sin\frac{\theta_\mathbf{k_2}}{2}\cos\frac{\theta_\mathbf{k_2}}{2}
e^{i\phi_\mathbf{k_2}}\nonumber\\&\times& \bigg{(}\frac{1}{E-\varepsilon_{\mathbf{k_1}}-\epsilon^+_{\mathbf{k_2}}}-
\frac{1}{E-\varepsilon_{\mathbf{k_1}}-\epsilon^-_{\mathbf{k_2}}}\bigg{)}.\nonumber
\end{eqnarray}
Here $\theta_{\mathbf{k}}$ and $\phi_{\mathbf{k}}$ stand for the polar angle and azimuthal angle of $\mathbf{k}$.

To solve Eq.(\ref{2-body eqn}), we define two auxiliary variables $f^{}_\sigma=g\sum_{\mathbf{p}}\Psi_\sigma(\mathbf{K-p,p}),\ \sigma=\uparrow,\downarrow$. After summing over $\mathbf{p}$ on both sides of Eq.(\ref{2-body eqn}), we obtain a closed equation of $f_\sigma$,
\begin{eqnarray}
f_\sigma=\frac g \Omega\sum_{\sigma',\mathbf{p}}G_{\sigma\sigma'}(\mathbf{p,K-p})
f_{\sigma'}.
\end{eqnarray}
We can calculate the summation explicitly in the above equation and write it in a compact matrix form,
\begin{eqnarray}
\left(
  \begin{array}{cc}
    D+Y\cos\theta & Y\sin\theta e^{-i\phi} \\
    Y\sin\theta e^{i\phi} &  D-Y\cos\theta\\
  \end{array}
\right)
\left(
  \begin{array}{c}
    f_\uparrow \\
    f_\downarrow \\
  \end{array}
\right)=0.\label{2-body matrix eqn}
\end{eqnarray}
Here $\theta$ and $\phi$ stand for $\theta_{\mathbf{K}}$ and $\phi_{\mathbf{K}}$. $D$ and $Y$ are two analytical functions of $E$, $a$, $|\mathbf{K}|$, and $\mu=M/m$, whose expressions are listed in Appendix A. The matrix Eq.(\ref{2-body matrix eqn}) has a nontrivial solution 
only when the following equation is satisfied:
\begin{eqnarray}
\det\left(
  \begin{array}{cc}
    D+Y\cos\theta & Y\sin\theta e^{-i\phi} \\
    Y\sin\theta e^{i\phi} &  D-Y\cos\theta\\
  \end{array}
\right)=D^2-Y^2=0,\label{2-body final}
\end{eqnarray}
from which we extract the two-body binding energy $E$. We can see that $E$ is independent of the direction of total momentum $\mathbf{K}$, which is the consequence of the SU(2) invariant interaction and the isotropic SO coupling.

\begin{figure}[t]
\includegraphics[width=3.4 in]
{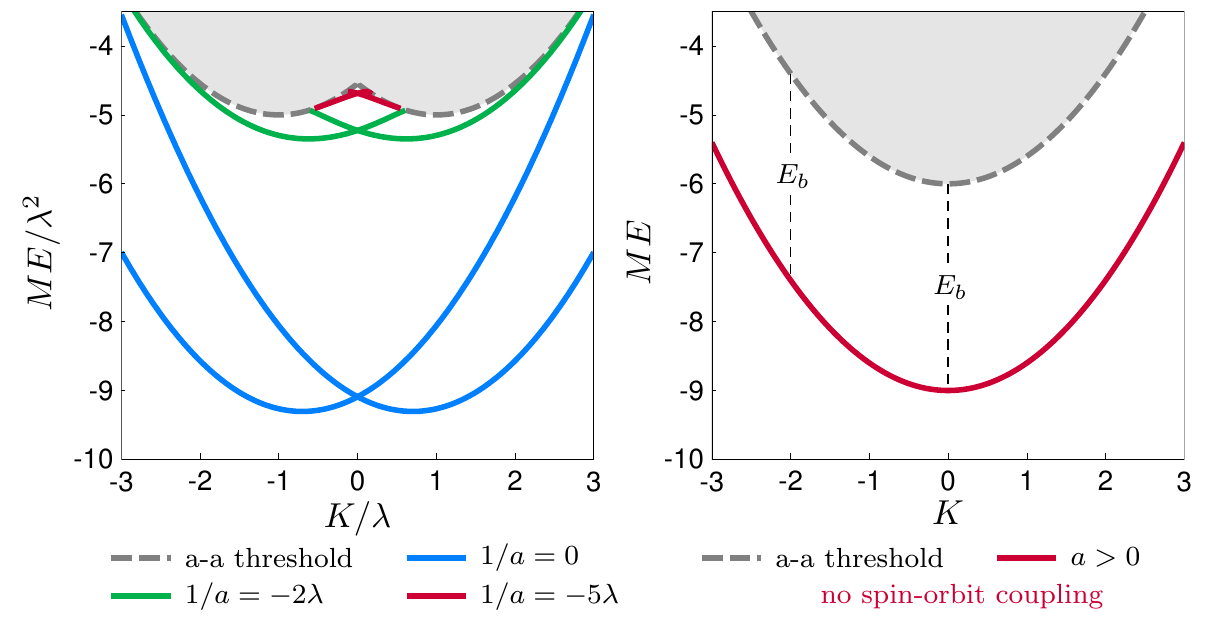}
\caption{(Color online). Two-body binding energy as a function of total momentum $\mathbf{K}$(in unit of SO coupling strength $\lambda$) for a SO coupled system(left panel) and a system without SO coupling(right panel). The mass ratio is set to be $\mu=1$. \label{fig1}}
\end{figure}

In Fig.\ref{fig1} we show the numerical results of the binding energies of a two-body system with or without SO coupling. We can find several interesting effects of SO coupling:

(i) In $\mathbf{K=0}$ subspace, the SO coupled system can support two degenerate bound states for arbitrary scattering length and mass ratio. While for a system without SO coupling, there is no two-body bound state for negative scattering length. This is because the SO coupling enhances the low-energy density of states, which makes the system behaves more like a one dimensional system. We note that similar effect has been found in other SO coupled two-body systems\cite{Vijay,Yu_Zhai,Cui,Galitski,Yu}.

(ii) While increasing the total momentum $|\mathbf{K}|$, some of the bound states will merge into the two-atom threshold and dissociate into two atoms. This phenomenon can be understood by considering the large $\mathbf{K}$ limit. Since the spin-orbit coupling only affects the single particle dispersion, while increasing the total momentum $\mathbf{K}$, the deviation of the dispersion becomes more and more negligible. At very large $\mathbf{K}$, the system behaves like a system without SO coupling. Consequently, for $a<0$ system, all the bound states would eventually merge into the two-atom threshold at some sufficiently large $|\mathbf{K}|$, while for $a>0$ system, the bound state would never dissociate, and the bound state energy at large $|\mathbf{K}|$ can be approximated by an asymptotic form $E(\mathbf{K})\sim\frac{K^2}{2(M+m)}-\frac{M+m}{2mMa^2}+O(\frac\lambda K)$.

(iii) Different from the system without SO coupling, in which the two-body bound state energy takes a simple form $E(\mathbf{K})=E(0)+\frac{K^2}{2(M+m)}$, the momentum dependence of two-body bound states in the SO coupled system is very different and cannot be written in a simple analytic form. This is caused by the SO coupling term which breaks the Galilean invariance of the Hamiltonian and couples the center of mass degree of freedom to the relative motion.

\section{\uppercase\expandafter{\romannumeral3}. three-body system --- Born-Oppenheimer approximation}
In the rest of this paper, we will focus on the three-body system with two identical fermions ($\alpha$) interacting with another atom $\beta$, and $\beta$ is subjected to an isotropic SO coupling. The three-body Hamiltonian reads:
\begin{eqnarray}
H_{\text{3b}}=\frac{\mathbf{p_1}^2}{2M}+\frac{\mathbf{p_2}^2}{2M}+\frac{\mathbf{p_3}^2}{2m}-\frac{\lambda\mathbf{p_3}\cdot\hat{\sigma}}{m}\nonumber\\
+g\delta(\mathbf{r_1-r_3})+g\delta(\mathbf{r_2-r_3}).\label{3-body H}
\end{eqnarray}


In order to get an intuitive picture of this three-body problem, we first use the Born-Oppenheimer approximation to analyze this system in the $M\gg m$ limit. In this limit, since the two $\alpha$-atoms are much heavier than the SO coupled $\beta$-atom, we can assume that the $\beta$-atom is moving around two fixed heavy atoms located at $\mathbf{R_1}$ and $\mathbf{R_2}$. Its wave function can be written as $\varphi_\sigma(\mathbf{r})$ which satisfies the following noninteracting Schr\"{o}dinger equation plus the Bethe Peierls boundary condition,
\begin{eqnarray}
\bigg{[}-\frac{\nabla_{\mathbf{r}}^2}{2m}+\frac{i\lambda\nabla_{\mathbf{r}}\cdot\hat{\sigma}}{m}\bigg{]}\varphi_\sigma(\mathbf{r})=\epsilon\varphi_\sigma(\mathbf{r}),\label{B.O.eq1}\\
\varphi_\sigma\propto(\frac{1}{|\mathbf{r-R}_i|}-\frac{1}{a}),\quad\text{as }|\mathbf{r-R}_i|\rightarrow0,\label{bp}
\end{eqnarray}
where $\mathbf{R}_i,\ i=1,2$ are the positions of two heavy $\alpha$-atoms and $\mathbf{r}$ refers to the position of $\beta$-atom.

After solving Eq.(\ref{B.O.eq1}), we can obtain the three-body binding energy through solving the Schr\"{o}dinger equation for two heavy atoms which now interact via the effective potential $\epsilon(\mathbf{R_1,R_2})$,
\begin{eqnarray}
\bigg{[}-\frac{\nabla_1^2}{2M}-\frac{\nabla_2^2}{2M}+\epsilon(\mathbf{R_1,R_2})\bigg{]}\psi(\mathbf{R_1,R_2})=E\psi(\mathbf{R_1,R_2}).\nonumber\\\label{B.O.eq2}
\end{eqnarray}

Because of the translational symmetry of the system, we have $\epsilon(\mathbf{R_1,R_2})=\epsilon(\mathbf{R_1-R_2})$. Without loss of generality, we assume that $\mathbf{R_1=0}$ and $\mathbf{R_2=R}$. It can be proved that the solution of Eq.(\ref{B.O.eq1}) can be expressed as
\begin{eqnarray}
\varphi_\sigma(\mathbf{r})=\sum_{\sigma'=\uparrow,\downarrow}f_{1,\sigma'}G^{(0)}_{\sigma,\sigma'}(\mathbf{r,0})+f_{2,\sigma'}G^{(0)}_{\sigma,\sigma'}(\mathbf{r,R}).
\end{eqnarray}
Here $f_{1,\sigma},f_{2,\sigma}$ are constants remain to be determined, $G^{(0)}_{\sigma,\sigma'}(\mathbf{r_1,r_2})=G^{(0)}_{\sigma,\sigma'}(\mathbf{r_1-r_2})$ is the Green's function in real space for a single SO coupled $\beta$-atom. After straightforward calculation, we find that
\begin{eqnarray}
G_{\uparrow\uparrow}(\mathbf{r})&=&G^*_{\downarrow\downarrow}(\mathbf{r})=A+iB\cos{\theta_\mathbf{r}},\\
G_{\uparrow\downarrow}(\mathbf{r})&=&-G^*_{\downarrow\uparrow}({\mathbf{r}})=iB\sin{\theta_\mathbf{r}}e^{-i\phi_\mathbf{r}}.
\end{eqnarray}
Here $\theta_\mathbf{r}$ and $\phi_\mathbf{r}$ stand for the polar angle and azimuthal angle of $\mathbf{r}$. $A$ and $B$ are two analytical functions of $|\mathbf{r}|$ and $\epsilon$, which are listed in Appendix A.

Then we apply the Bethe Peierls boundary condition (\ref{bp}) and find following matrix equation for $f_{i,\sigma}$,
\begin{eqnarray}
C_M
\left(
  \begin{array}{c}
    f_{1\downarrow} \\
    f_{1\uparrow} \\
    f_{2\downarrow} \\
    f_{2\uparrow} \\
  \end{array}
\right)=0.
\end{eqnarray}
The coefficient matrix can be found in Appendix A.

The existence of a set of nontrivial solution requires the determinant of the coefficient matrix vanishes, we find
\begin{eqnarray}
\det C_M=A^2+B^2-(\frac{1}{a}-\sqrt{2m|\epsilon|}+\frac{\lambda^2}{\sqrt{2m|\epsilon}|})^2=0.\nonumber\\\label{B.O. algebraic}
\end{eqnarray}
Similar to Eq.(\ref{2-body final}), the effective potential $\epsilon(R)$ does not depend on the direction of $\mathbf{R}$.

\begin{figure}
\includegraphics[width=3.5 in]
{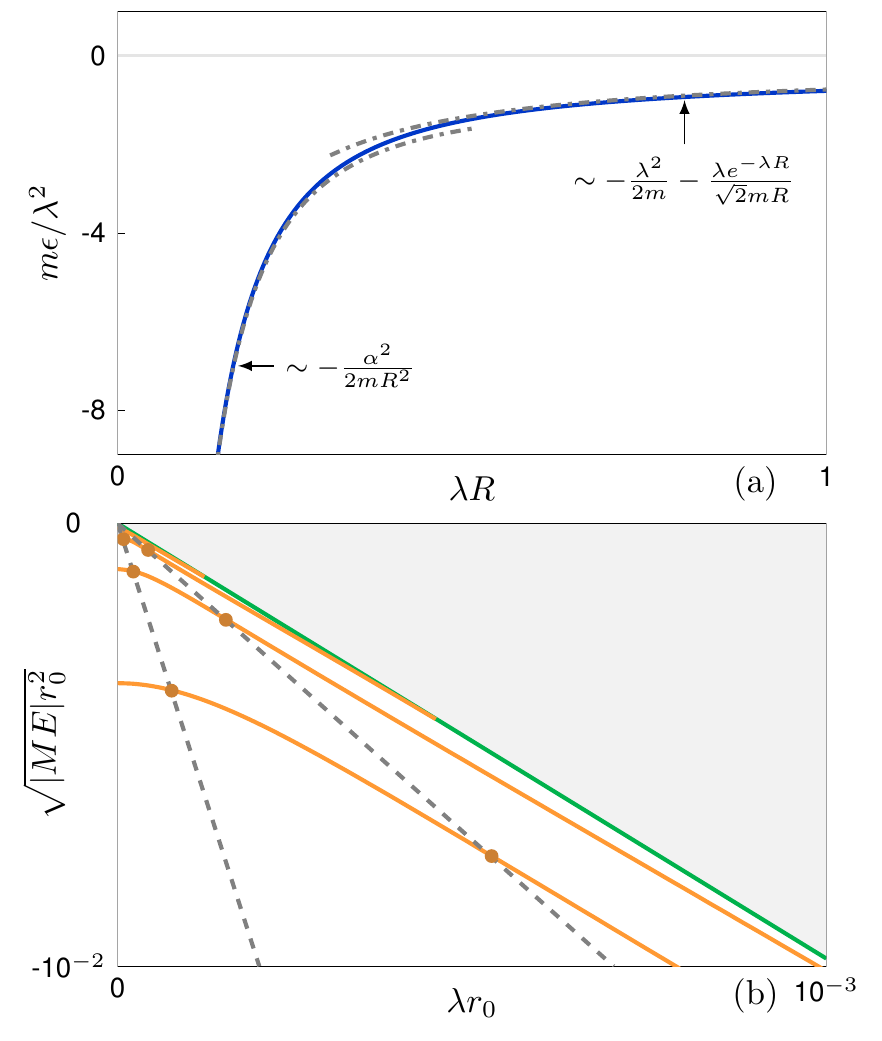}
\caption{(Color online). (a): Effective potential for two heavy atoms for scattering length $a=\infty$.
(b): Three-body bound state energies as a function of SO coupling strength $\lambda$ (solid orange). Threshold energy (solid green). The mass ratio is chosen to be $50$.\label{fig2}}
\end{figure}

Before we use Eq.(\ref{B.O. algebraic}) to solve the effective potential $\epsilon(R)$ directly, we shall analyze its asymptotic behavior. This can help us understand the role of SO coupling in the bound state problem.

At small $R$ limit, we find
\begin{eqnarray}
\epsilon(R)\sim-\frac{\alpha^2}{2mR^2}, \label{short-range}
\end{eqnarray}
where $\alpha\simeq0.517643$ is the solution of the equation $x=e^{-x}$. Note that the asymptotic form is exactly the effective potential of a system without SO coupling. This connection is natural, because in the short range limit, those parameters which have the same dimension of length can be considered as effectively divergent, which makes $a\sim\infty$ and $\lambda\sim0$.

It is known that the short range $-\frac{1}{R^2}$ potential is the origin of the Efimov physics\cite{Efimov,Braaten}. For a pure $-\frac{\alpha^2}{2mR^2}$ potential, if the mass ratio is beyond a critical value $\mu>{9}/{2\alpha^2}\simeq14.0$, the system can support a series of Efimov states with discrete scaling. The scaling factor is $e^{\pi/s_0}$, where $s_0=\sqrt{\alpha^2\mu-9/2}$.

In our SO coupled case, the short range $-\frac{1}{R^2}$ behavior brings two effects. First, the fact that the short range potential remains unchanged in SO coupled system means that the high energy physics
is insensitive to the SO coupling. More specifically, for those deeply bound Efimov states,
their binding energies (with $E_b\gg\frac{\lambda^2}{2M}$) are hardly affected by SO coupling. Second, the $-\frac{1}{R^2}$ potential implies that our system may still exhibit some discrete scaling, even though we have introduced a length scale $\lambda$ into the system.

The asymptotic behavior of $\epsilon(R)$ at long range limit is also very important. In this limit, we can ignore the other heavy particle when the light atom is closed to $\mathbf{R_1}$. Then the wave function $\varphi_\sigma$ can be well approximated by the linear combination of two two-body wave functions,
\begin{eqnarray}
\varphi_\sigma(\mathbf{r})\sim C_1\phi_\sigma(\mathbf{r-R_1})+C_2\phi_\sigma(\mathbf{r-R_2}), \label{asymptotic}
\end{eqnarray}
where $\phi_\sigma(\mathbf{r-R}_i)$ is the two-body bound state wave function when there is only one heavy atom located at $\mathbf{R}_i$. The asymptotic behavior of $\varphi_\sigma$ should reflect the inversion symmetry of $\varphi_\sigma$ itself, which means the R.H.S. of Eq.(\ref{asymptotic}) should remain unchanged(up to a global phase $e^{i\theta}$) after we exchange $\mathbf{R_1}$ and $\mathbf{R_2}$. This leads to $C_1=C_2$ or $C_1=-C_2$, and it is easy to check that the former always has the lower energy.

Therefore, the asymptotic form of $\epsilon(R)$ in the large $R$ limit should be equal to the two-body binding energy plus a correction due to the overlap integral of $\phi_\sigma(\mathbf{r-R}_1)$ and $\phi_\sigma(\mathbf{r-R}_2)$. Indeed, by taking $R\rightarrow\infty$ limit of Eq.(\ref{B.O. algebraic}), we find
\begin{eqnarray}
\epsilon(R)\sim-E_{2b}-\frac{e^{-\kappa_0R}}{R}\bigg{[}\frac{\kappa_0^2a\sqrt{1+\lambda/\kappa_0}}{m(2\kappa_0a-1)}\bigg{]}, \label{long-range}
\end{eqnarray}
where $\kappa_0=\frac{1}{2a}(1+\sqrt{1+4\lambda^2a^2})$, and $E_{2b}=\frac{\kappa_0^2}{2m}$ is the two-body binding energy.

This fast decay behavior as shown in Eq.(\ref{long-range}) will have significant effects on shallow Efimov states with binding energies $E_b\lesssim\frac{\lambda^2}{2M}$, as they have very extended wave functions and thus are more sensitive to the long-range part of $\epsilon(R)$. Since the long-range effective potential decays much faster than the pure $-\frac{1}{R^2}$ potential, these shallow bound states are expected to be less favored in the presence of SO coupling.


In Fig.2(a), we plot $\epsilon(R)$ as a function of $R$ at $a=\infty$ by numerically solving Eq.(\ref{B.O. algebraic}), and also verify its short-range and long-range behaviors as shown by Eq.(\ref{short-range}) and Eq.(\ref{long-range}). In Fig.2(b), we show the three-body spectrum as increasing the SO coupling strength $\lambda$, which is obtained by applying a hard-core boundary condition $\psi|_{R=r_0}=0$ to Eq.(\ref{B.O.eq2}). We can see that the binding energies of those deeply bound states are insensitive to the SO coupling for small $\lambda$. While if we keep increasing $\lambda$, all bound states will eventually merge into the atom-dimer continuum and disappear. This behavior verifies our analyses based on the asymptotic behaviors of $\epsilon(R)$.

Moreover, we find an interesting scaling behavior of these three-body bound states. We checked the ratio of two successive bound state energies along the gray dashed line in Fig.\ref{fig2}(b). It shows that the energy ratios follow a discrete scaling law,
\begin{eqnarray}
\frac{E_{n+1}(\lambda,a)}{E_n(e^{\pi/s_0}\lambda,e^{-\pi/s_0}a)}\backsimeq e^{-2\pi/s_0}.\label{discrete scaling}
\end{eqnarray}
This discrete scaling behavior is also due to the short-range $-\frac{1}{R^2}$ effective potential. For an arbitrary bound state wave function $\psi(\mathbf{R})$, we can do following scaling transformation,
\begin{eqnarray}
\mathbf{R}\rightarrow e^{-\pi/s_0}\mathbf{R},&\quad& a\rightarrow e^{-\pi/s_0}a,\nonumber\\
\lambda\rightarrow e^{\pi/s_0}\lambda,&\quad& E\rightarrow e^{2\pi/s_0}E,\label{scaling transformation}
\end{eqnarray}
where $s_0=\sqrt{\alpha^2\mu-9/2}$. After such transformation, the wave function still satisfies corresponding Schr\"{o}dinger equation under the same hard-core boundary condition. 
The modified discrete scaling law as Eq.(\ref{discrete scaling}) can thus be applied to the SO coupled system. 

In the end of this section, we conclude several important effects of SO coupling on the Efimov physics in this system.

(i) SO coupling will not change the critical mass ratio to support Efimov states.

(ii) As increasing the SO coupling, the Efimov states will merge into the atom-dimer continuum and disappear.

(iii) In the presence of SO coupling, the system exhibits a discrete scaling behavior like Eq.(\ref{discrete scaling}). Moreover, the scaling ratio is identical to that without SO coupling.

Although these properties are based on the Born-Oppenheimer approximation that is valid only for very large mass ratio, the calculation in next section will show that all the above properties hold exactly for arbitrary mass ratio.

\section{\uppercase\expandafter{\romannumeral4}. three-body system --- Exact solution}

In this section, we exactly solve the three-body bound state problem. Subsection \uppercase\expandafter{\romannumeral4} A and \uppercase\expandafter{\romannumeral4}  B are respectively for the discussion of Efimov trimer and universal trimer states.

Similar to the two-body problem, we shall define a three-body threshold energy at first. Generally speaking, a three-body bound state can either dissociate into three free atoms or one free atom plus one two-body bound state(dimer). These two channels give two corresponding thresholds, three-atom(aaa) threshold and atom-dimer(a-d) threshold which are defined as following,
\begin{eqnarray}
E_{\text{aaa}}(\mathbf{K})&=&\min_{\mathbf{k_1+k_2+k_3=K},\pm}\bigg{(}\varepsilon_{\mathbf{k_1}}+\varepsilon_{\mathbf{k_2}}+\epsilon^\pm_{\mathbf{k_3}}\bigg{)},\\
E_{\text{a-d}}(\mathbf{K})&=&\min_{\mathbf{k+p=K}}\bigg{(}\varepsilon_{\mathbf{k}}+E_2(\mathbf{p})\bigg{)}.
\end{eqnarray}
Here $\varepsilon_\mathbf{k}=k^2/2M$, $\epsilon^\pm_\mathbf{k}=(k^2\pm2|k|\lambda)/2m$ are the dispersions for $\alpha$-atom and $\beta$-atom. $E_2(\mathbf{p})$ is the two-body bound state energy calculated in Section \uppercase\expandafter{\romannumeral2}. Since the total momentum $\mathbf{K}$ is always a good quantum number, both thresholds are defined as a function of $\mathbf{K}$.

The three-body threshold $E_{\text{th}}$ is the minimum of both three-atom threshold and atom-dimer threshold,
\begin{eqnarray}
E_{\text{th}}(\mathbf{K})=\min\bigg{\{}E_{\text{aaa}}(\mathbf{K}),E_{\text{a-d}}(\mathbf{K})\bigg{\}}.
\end{eqnarray}

\begin{figure}
\includegraphics[width=3 in]
{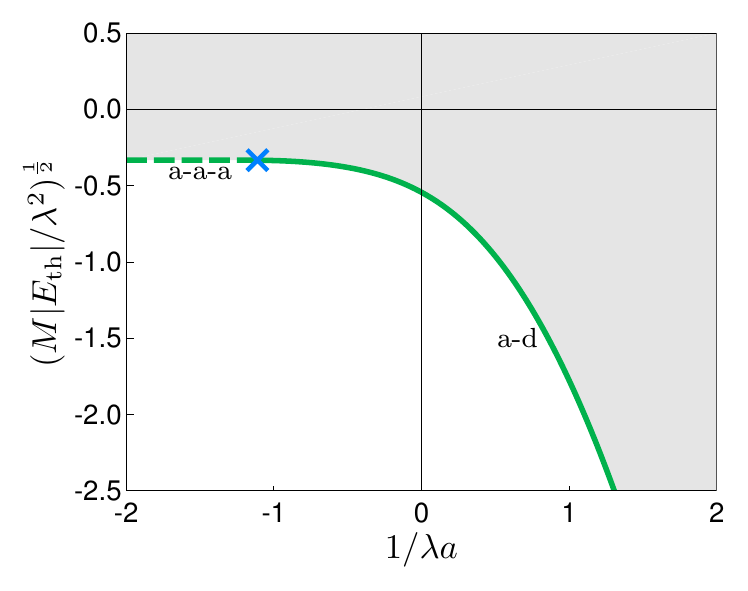}
\caption{(Color online). Three-body threshold energy $E_{\text{th}}(0)$ as a function of $1/\lambda a$. (mass ratio $M/m=\mu=1$)\label{fig3}}
\end{figure}

In Fig.\ref{fig3}, we plot the three-body threshold as a function of $1/\lambda a$ in $\mathbf{K=0}$ subspace. Different from the case without SO coupling, the atom-dimer threshold becomes larger than the three-atom threshold at some critical point. This is due to the complex momentum dependence of the two-body binding energy.

We restrict the calculation in a fixed total momentum subspace. The Lippmann-Schwinger equation is
\begin{eqnarray}
\Psi_\sigma(\mathbf{k_1,k_2})&=
&\frac{g}{\Omega}\sum_{\mathbf{q},\sigma'}G^{(0)}_{\sigma\sigma'}(\mathbf{k_1,k_2})\nonumber
\\&\times&\bigg{[}\Psi_{\sigma'}(\mathbf{q,k_2})+\Psi_{\sigma'}(\mathbf{k_1,q})\bigg{]}.\label{three_body_eqn}
\end{eqnarray}
Here $\mathbf{k_1,k_2}$ are the momenta of two $\alpha$-atoms. We omit the momentum of the $\beta$-atom in the equation since the total momentum $\mathbf{K}$ is always a conserved quantity. So $\Psi_\sigma(\mathbf{k_1,k_2})$ is actually a short form of $\Psi_\sigma(\mathbf{k_1,k_2,K-k_1-k_2})$. $G^{(0)}_{\sigma\sigma'}$ is the Green's function of three noninteracting particles,
\begin{eqnarray}
G_{\uparrow\uparrow}&=&\frac{\cos^2\frac{\theta_\mathbf{k_3}}{2}}
{E-\varepsilon_{\mathbf{k_1}}-\varepsilon_{\mathbf{k_2}}-\epsilon^+_{\mathbf{k_3}}}+
\frac{\sin^2\frac{\theta_\mathbf{k_3}}{2}}
{E-\varepsilon_{\mathbf{k_1}}-\varepsilon_{\mathbf{k_2}}-\epsilon^-_{\mathbf{k_3}}},\nonumber\\
G_{\downarrow\downarrow}&=&\frac{\sin^2\frac{\theta_\mathbf{k_3}}{2}}
{E-\varepsilon_{\mathbf{k_1}}-\varepsilon_{\mathbf{k_2}}-\epsilon^+_{\mathbf{k_3}}}+
\frac{\cos^2\frac{\theta_\mathbf{k_3}}{2}}
{E-\varepsilon_{\mathbf{k_1}}-\varepsilon_{\mathbf{k_2}}-\epsilon^-_{\mathbf{k_3}}}, \nonumber\\
G_{\downarrow\uparrow}&=&G^*_{\uparrow\downarrow}=\sin\frac{\theta_\mathbf{k_3}}{2}\cos\frac{\theta_\mathbf{k_3}}{2}
e^{i\phi_\mathbf{k_3}}\\
&\times&\bigg{(}\frac{1}{E-\varepsilon_{\mathbf{k_1}}-\varepsilon_{\mathbf{k_2}}-\epsilon^+_{\mathbf{k_3}}}-
\frac{1}{E-\varepsilon_{\mathbf{k_1}}-\varepsilon_{\mathbf{k_2}}-\epsilon^-_{\mathbf{k_3}}}\bigg{)}.\nonumber
\end{eqnarray}

To solve Eq.(\ref{three_body_eqn}), we define an auxiliary function $f_{\sigma}(\mathbf{p})$ as
\begin{eqnarray}
f_\sigma(\mathbf{p})=g\sum_{\mathbf{q}}\Psi_\sigma(\mathbf{q,K-p})
=- g\sum_{\mathbf{q}}\Psi_\sigma(\mathbf{K-p,q}).\nonumber\\
\end{eqnarray}
Using this definition, we can simplify Eq.(\ref{three_body_eqn}) into
\begin{eqnarray}
f_\sigma(\mathbf{k})=\frac{g}{\Omega}\sum_{\sigma',\mathbf{p}}G^{(0)}_{\sigma\sigma'}(\mathbf{p,K-k})
\bigg{[}f_{\sigma'}(\mathbf{k})- f_{\sigma'}(\mathbf{K-p})\bigg{]}.\nonumber\\\label{three_body_eqn_simplified}
\end{eqnarray}

The linear integral Eq.(\ref{three_body_eqn_simplified}) has one trivial solution which is $f_\sigma=0$. For some special $E=E_3(\mathbf{K})$ the equation has nonzero solution of $f_\sigma$, this gives the energies of three-body bound states. 
Next we show that the symmetry consideration can simplify the problem to a great extent.

There are two good quantum numbers of the three-body Hamiltonian, the total momentum $\mathbf{P=p_1+p_2+p_3}$ and the total angular momentum $\mathbf{J=L+s}$. They correspond to the spatial translation operation and a simultaneously rotation in real and spin space. We find following commutation relation of $\mathbf{P}$ and $\mathbf{J}$,
\begin{eqnarray}
[J_i,J_j]=i\epsilon_{ijk}J_k,\quad[P_i,P_j]=0,\quad[P_i,J_j]=i\epsilon_{ijk}P_k.\nonumber\\\label{algebra}
\end{eqnarray}
Here the subindices $i,j,k=1,2,3$ denote different components of $\mathbf{P}$ and $\mathbf{J}$, and $\epsilon_{ijk}$ is the common Levi-Civita symbol.

The commutation relation(\ref{algebra}) is the algebra of the special Euclidean group SE(3), which is related to the kinematics of a rigid body in three dimension. It has been proved that, there are two independent Casimir invariants for SE(3), which are $\mathbf{P^2}$ and $\mathbf{P\cdot J}$\cite{group1}. Since we have already used the conservation of total momentum $\mathbf{P}$, the only nontrivial Casimir is then $\mathbf{P\cdot J}$. In Appendix B, we show that the use of this Casimir can help integrate out one fold of the integral in Eq.(\ref{three_body_eqn_simplified}) and simplify it into a two dimensional integral equation.

Moreover, the commutation relation(\ref{algebra}) suggests that the angular momentum $\mathbf{J}$ actually commutes with the total momentum $\mathbf{P}$ in $\mathbf{P=0}$ subspace. Therefore, in this subspace, we can use the good quantum number $\mathbf{J}$. In Appendix C, we show that for a bound state with angular momentum $(J,J_z)=(j+\frac{1}{2},m+\frac{1}{2})$, $f_\sigma(\mathbf{k})$ should take the form,
\begin{eqnarray}
f_\uparrow(\mathbf{k})&=&\sqrt{\frac{j+m+1}{2j+1}}f_0Y_j^m-\sqrt{\frac{j-m+1}{2j+3}}f_1Y_{j+1}^m,\nonumber\\
f_\downarrow(\mathbf{k})&=&\sqrt{\frac{j-m}{2j+1}}f_0Y_j^{m+1}+\sqrt{\frac{j+m+2}{2j+3}}f_1Y_{j+1}^{m+1},\nonumber\\\label{ansatz}
\end{eqnarray}
where $f_0$ and $f_1$ are two functions only depending on the magnitude of $\mathbf{k}$ and $Y_j^m$ is short for $Y_j^m(\Omega_\mathbf{k})$.

After substituting this ansatz into Eq.(\ref{three_body_eqn_simplified}), we obtain two coupled one dimensional integral equations which can be written in a compact form,
\begin{eqnarray}
Z(k)\left(
      \begin{array}{c}
        f_0(k) \\
        f_1(k) \\
      \end{array}
    \right)
=\int_0^\Lambda dpK_j(k,p)
\left(
  \begin{array}{c}
    f_0(p) \\
    f_1(p) \\
  \end{array}
\right).\label{final integral}
\end{eqnarray}
The high energy cutoff $\Lambda$ is equivalent to imposing a short-range three-body boundary condition as we did in Section \uppercase\expandafter{\romannumeral3}. Both $Z$ and $K_j$ are two-by-two matrices, whose elements are shown in Appendix C. We can see that the bound states with different quantum numbers $J_z$ are degenerate, which is also a consequence of the SE(3) symmetry.

\subsection{A. Efimov trimer}

Since we know that the emergence of the Efimov physics comes from the high momentum part of the coefficient matrices $Z$ and $K_J$. While in the high momentum region, where $p_3\gg\lambda$, the dispersion $\epsilon^{\pm}_{\mathbf{p_3}}$ becomes closer to a normal parabolic form and the SO coupling effect can be neglected. Therefore, the SO coupling will not change the critical mass ratio for Efimov trimers:
\begin{eqnarray}
\mu_{\text{\tiny{Efimov}}}(\lambda)=\mu_{\text{\tiny{Efimov}}}(0)=13.606...
\end{eqnarray}

\begin{figure}
\includegraphics[width=3 in]
{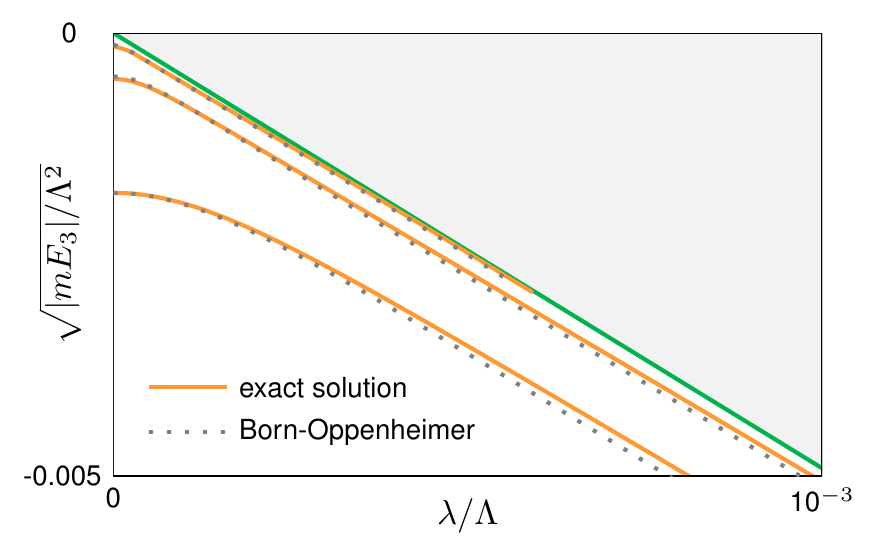}
\caption{(Color online). Three-body bound state($J=1/2$) energies as a function of SO coupling strength $\lambda$ at $a=\infty$. Solid orange: Exact solution. Dotted gray: Born-Oppenheimer approximation. Solid green: Threshold energy. The mass ratio is chosen to be $50$.\label{fig4}}
\end{figure}

\begin{figure}
\includegraphics[width=3.4 in]
{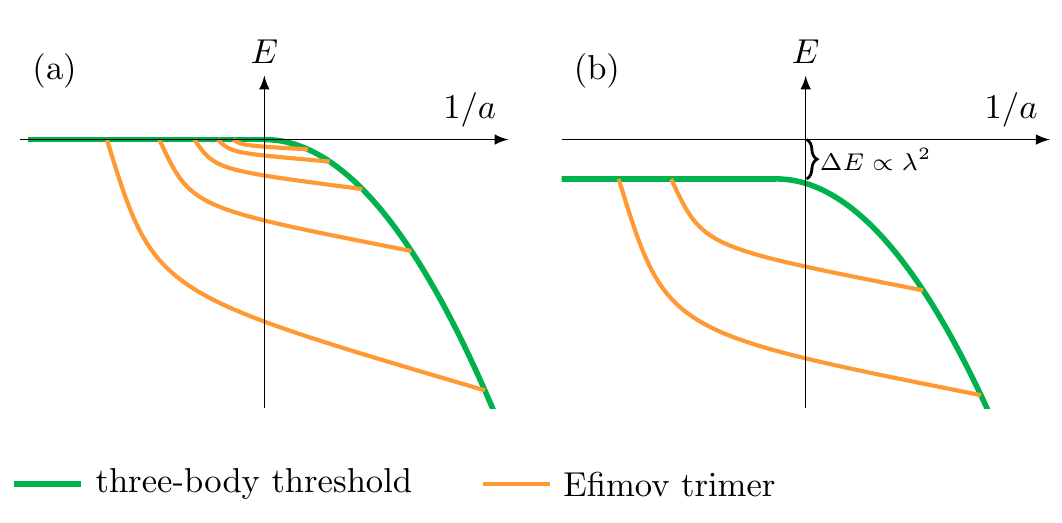}
\caption{(Color online). Schematic plots of three-body binding energy. (a): the system without coupling. (b): SO coupled system.\label{fig5}}
\end{figure}

In Fig.\ref{fig4}, we plot the trimer energies in the lowest angular momentum channel ($J=1/2$) as a function of SO coupling strength $\lambda$ at two-body resonance. We find that many shallow bound states will merge into the atom-dimer threshold as we increasing $\lambda$, which confirms our previous conclusion based on Born-Oppenheimer approximation. In Fig.\ref{fig5}, we further give a schematic plot of how the Efimov states changes after switching on the SO coupling for finite scattering lengths.
Moreover, we also numerically verify the discrete scaling behavior in Eq.(\ref{discrete scaling}), where $s_0$ is exactly the same value as that without SO coupling.

\subsection{B. Universal trimer}

Besides the Efimov trimers, it is known that there exists another kind of three-body bound states if the mass ratio is within a region smaller than $\mu_c\simeq13.606...$\cite{KM,Ueda,SCZ}. These states are called as universal trimers as their binding energies do not depend on the details of short-range interaction potentials (equivalent to high-energy cutoff in zero-range model). 
As a result, the universal trimers can have following continuous scaling behavior,
\begin{eqnarray}
a\rightarrow \alpha^{-1} a,\quad\lambda\rightarrow \alpha\lambda,\quad E\rightarrow \alpha^{2}E,\label{continuous scaling}
\end{eqnarray}
which can be proved by a simple dimensional analysis. Note that here $\alpha$ can be an arbitrary positive number.


In Ref.\cite{SCZ}, we have calculated the lowest trimer energy in $J=3/2$ channel for different high energy cutoff $\Lambda$. It turns out that the binding energies for different $\Lambda$ coincide with each other for mass ratio smaller than $13.6$. Once the mass ratio exceeds $13.606$, these universal trimers become the lowest Efimov bound states whose energies will depend on the high energy cutoff $\Lambda$\cite{Ueda}. This proves that these three-body bound states are universal trimers for mass ratio below $\mu_c$.

\begin{figure}
\includegraphics[width=3.5 in]
{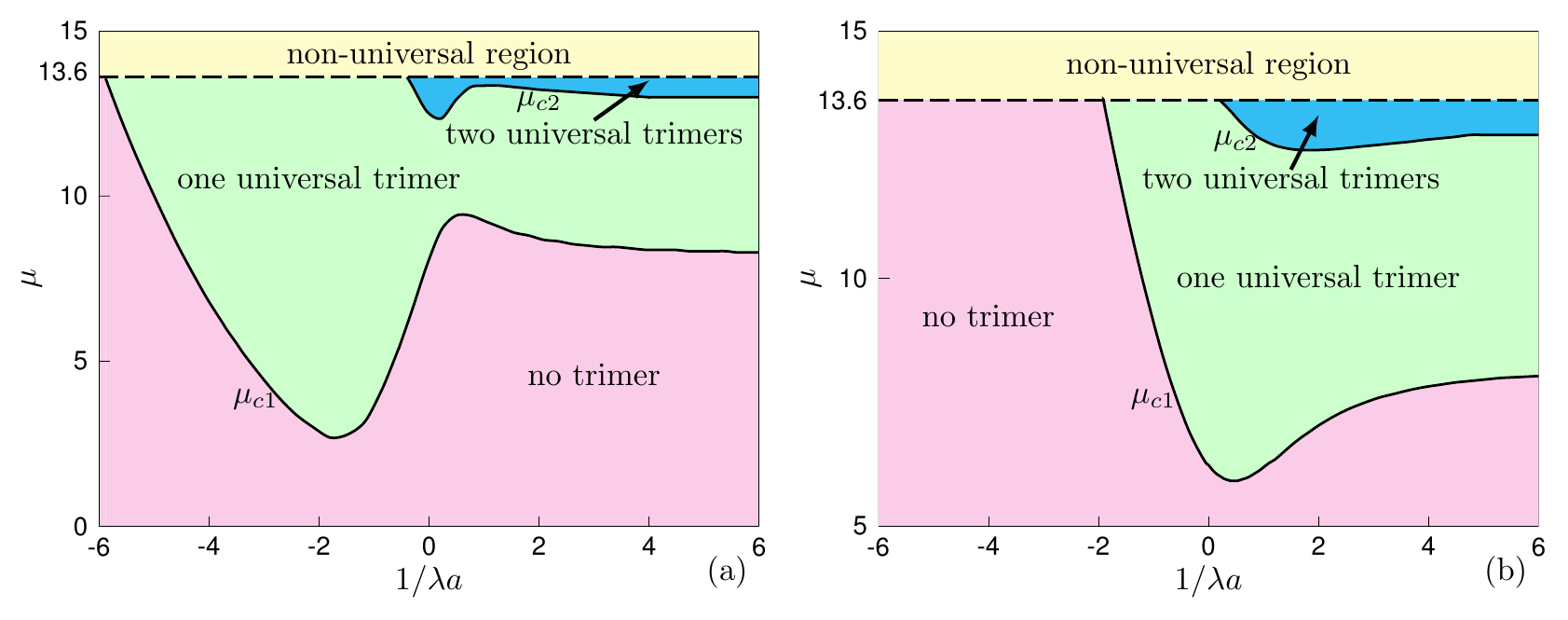}
\caption{(Color online). The critical mass ratios for universal trimers as functions of $1/\lambda a$. (a): $J=1/2$ channel. (b): $J=3/2$ channel. The plot of $J=3/2$ channel are from reference\cite{SCZ}\label{fig6}}
\end{figure}

In Fig.\ref{fig6}, we plot the critical mass ratios of two lowest angular momentum channels ($J=1/2,\ 3/2$) as functions of $1/\lambda a$. The phase diagrams are separated into several different regions, where there is one, two or no universal trimers. It is found that in $1/\lambda a\rightarrow +\infty$ limit, the critical mass ratios in different channels approach to same numbers,
\begin{eqnarray}
  \begin{array}{c}
    \mu_{c1}(J=\frac12)=\mu_{c1}(J=\frac32)\simeq8.172 \\
    {}\\
    \mu_{c2}(J=\frac12)=\mu_{c2}(J=\frac32)\simeq12.92 \\
  \end{array}
\quad,1/\lambda a\rightarrow+\infty.
\end{eqnarray}
These two numbers are exactly the same critical mass ratios to support the so called Kartavtsev-Malykh trimers in the system without SO coupling\cite{KM}. Since the $1/\lambda a\rightarrow +\infty$ limit can be interpreted as fixing $a$ (as a positive constant ) and reducing $\lambda$ to zero. This means the $1/\lambda a\rightarrow +\infty$ can be considered as no SO coupling limit. In this sense, these universal trimers can also be considered to be the bound states continuously connected with the Kartavtsev-Malykh trimers.

However, we shall still point out several differences between our universal trimers and Kartavtsev-Malykh trimers. First, the Kartavtsev-Malykh trimers can only exist in $a>0$ region, while in SO coupled system, the universal trimers can live in both $a>0$ and $a<0$ sides of the Feshbach resonance. Moreover, the critical mass ratios for Kartavtsev-Malykh trimers are fixed and cannot be tuned by any other parameters. While in our system, the critical mass ratios can be tuned by scattering length $a$ and SO coupling strength $\lambda$. This leads to a very rich and interesting phase diagram as shown in Fig.\ref{fig6}. We also find that the minimum critical mass ratio to support the bound states is lowered because of SO coupling. In $J=1/2$ channel, the minimum ratio is approximately $2.68$ and in $J=3/2$ channel, it is approximately $5.92$.

We note that SO coupling has very different effects on Efimov trimers and universal trimers. 
While the Efimov trimers are less favored by the SO coupling, the universal trimer are more favored and can be induced in a broader parameter regime (in terms of mass ratios and scattering lengths) compared to that without SO coupling. This makes the universal trimers much easier to be detected in atomic mixtures like Li-K-K system.

\section{\uppercase\expandafter{\romannumeral5}. summary}

We have studied the bound state properties of two-body and three-body systems when one of the atoms is subjected to an isotropic SO coupling. Our main results are summarized as follows.

1) 
We find that the two-body binding energy shows a very sensitive dependence on the total momentum.
In $\mathbf{K}=0$ subspace, the system can always support two-body bound states for arbitrary weak interactions.

2) For three-body system, the Born-Oppenheimer approximation shows that the three-body potential at resonance exponentially decays with the distance of two heavy fermions. As a result, the formation of three-body bound state (Efimov trimer) is less favored than that without SO coupling, and as increasing the SO coupling strength the Efimov trimers merge into the atom-dimer threshold.
Moreover, we show that the SO coupling does not change the critical mass ratio (13.606) for Efimov physics, while the Efimov trimers obey a new kind of discrete scaling law in the presence of SO coupling. These results are all confirmed by the exact solutions of three-body problem.

3) The presence of SO coupling can induce universal trimers at negative s-wave scattering length side and with smaller mass ratio than those without SO coupling (the Kartavtsev-Malykh trimers). We find the minimal mass ratios for the occurrence of universal trimers in two lowest angular momentum channels are respectively $2.68$ and $5.92$, which could be accessible by atomic mixtures in cold atoms experiments.

\textbf{Acknowledgements}: This work is supported by Tsinghua University Initiative Scientific Research Program, NSFC under Grant No. 11174176, 11104158, 11374177 and NKBRSFC under Grant No. 2011CB921500.



\begin{widetext}

\vspace{0.05in}

\section{appendix A}
We list the analytical functions $D$, $Y$, $A$ and $B$ here,
\begin{eqnarray}
D&=&-\frac{1}{2\pi(1+\mu)}\bigg{[}\frac{1}{a}-\nonumber\\
&&\frac{(K+\mu\lambda)\sqrt{\mu(K^2-2K\lambda-\mu\lambda^2)-2(1+\mu)2ME}+
(K-\mu\lambda)\sqrt{\mu(K^2+2K\lambda-\mu\lambda^2)-2(1+\mu)ME}}{2K(1+\mu)}\bigg{]},\nonumber\\
Y&=&-\frac{1}{12\pi(1+\mu)^2K^2}
\bigg{\{}\big{[}\mu(K^2+K\lambda-\mu\lambda^2)+3K^2-2(1+\mu)ME\big{]}\sqrt{\mu(K^2-2K\lambda-\mu\lambda^2)-2(1+\mu)2ME}\\ \nonumber
&&-\big{[}\mu(K^2-K\lambda-\mu\lambda^2)+3K^2-2(1+\mu)ME\big{]}\sqrt{\mu(K^2+2K\lambda-\mu\lambda^2)-2(1+\mu)ME}
\bigg{\}},\nonumber\\
A&=&\frac{e^{-\kappa R}}{R}\bigg{(}\cos\lambda R+\frac{\lambda}{\kappa}\sin\lambda R\bigg{)},\qquad B=\frac{e^{-\kappa R}}{R}\bigg{(}\frac{\lambda}{\kappa}\cos\lambda R-\sin\lambda R+\frac{\sin\lambda R}{\kappa}\bigg{)},\qquad\text{where } \kappa=\sqrt{2m|\epsilon|}.\nonumber
\end{eqnarray}

The coefficient matrix $C_M$ is
\begin{eqnarray}
C_M=
\left(
  \begin{array}{cccc}
   C & 0 & A-iB\cos{\theta} & -iB\sin{\theta}e^{-i\phi} \\
  A+iB\cos{\theta} & iB\sin{\theta}e^{-i\phi} & C & 0 \\
 0 & C & -iB\sin{\theta}e^{i\phi} & A+iB\cos{\theta} \\
iB\sin{\theta}e^{i\phi} & A+iB\cos{\theta} & 0 & C \\
\end{array}
\right),\nonumber
\end{eqnarray}
where $C=\frac{1}{a}-\kappa+\frac{\lambda^2}{\kappa}$.

\section{Appendix B}

In this appendix, we will simplify the three-body equation (\ref{three_body_eqn_simplified}) to a two-dimensional integral equation for a generally finite total momentum $\mathbf{K}$. The simplification can also be achieved by using the SE(3) symmetry of the Hamiltonian.

Since the binding energy is independent of the direction of total momentum $\mathbf{K}$, we assume $\mathbf{K}$ is along $z$ axis in the following calculation. We can write down Eq.(\ref{three_body_eqn_simplified}) explicity,
\begin{eqnarray}
Z_\uparrow(\mathbf{K,k})f_\uparrow(\mathbf{k})+S_\downarrow(K,\mathbf{k})f_\downarrow(\mathbf{k})&=&\int\frac{d^3\mathbf{p}}{(2\pi)^3}
P_\uparrow(\mathbf{K,k,p})f_\uparrow(\mathbf{p})+Q_\downarrow(\mathbf{K,k,p})f_\downarrow(\mathbf{p}),\nonumber\\
Z_\downarrow(\mathbf{K,k})f_\downarrow(\mathbf{k})+S_\uparrow(K,\mathbf{k})f_\uparrow(\mathbf{k})&=&\int\frac{d^3\mathbf{p}}{(2\pi)^3}
P_\downarrow(\mathbf{K,k,p})f_\downarrow(\mathbf{p})+Q_\uparrow(\mathbf{K,k,p})f_\uparrow(\mathbf{p}).\label{Ab1}
\end{eqnarray}
Here $Z_\updownarrow,S_\updownarrow,P_\updownarrow,Q_\updownarrow$ are defined as following,
\begin{eqnarray}
Z_\updownarrow(\mathbf{K,k})&=&
\int\frac{d^3\mathbf{p}}{(2\pi)^3}
\bigg{[}\frac{1\pm\cos\theta_{\mathbf{p}}}{2ME-(\mathbf{K-k})^2-(\mathbf{k-p})^2-\mu(p+\lambda)^2}+
\frac{1\mp\cos\theta_{\mathbf{p}}}{2ME-(\mathbf{K-k})^2-(\mathbf{k-p})^2-\mu(p-\lambda)^2}\bigg{]}-\frac{1}{g}\nonumber\\
S_\updownarrow(\mathbf{K,k})&=&
\int\frac{d^3\mathbf{p}}{(2\pi)^3}
\bigg{[}\frac{\sin\theta_{\mathbf{p}}e^{\pm i\phi_{\mathbf{p}}}}{2ME-(\mathbf{K-k})^2-(\mathbf{k-p})^2-\mu(p+\lambda)^2}-
\frac{\sin\theta_{\mathbf{p}}e^{\pm i\phi_{\mathbf{p}}}}{2ME-(\mathbf{K-k})^2-(\mathbf{k-p})^2-\mu(p-\lambda)^2}\bigg{]},\nonumber\\
P_\updownarrow(\mathbf{K,k,p})&=&
\frac{1\pm\cos\theta_{\mathbf{q}}}{2ME-(\mathbf{K-k})^2-(\mathbf{K-p})^2-\mu(|\mathbf{q}|+\lambda)^2}
+\frac{1\mp\cos\theta_{\mathbf{q}}}{2ME-(\mathbf{K-k})^2-(\mathbf{K-p})^2-\mu(|\mathbf{q}|-\lambda)^2},\nonumber\\
Q_\updownarrow(\mathbf{K,k,p})&=&
\frac{\sin\theta_{\mathbf{q}}e^{\mp i\phi_{\mathbf{q}}}}{2ME-(\mathbf{K-k})^2-(\mathbf{K-p})^2-\mu(|\mathbf{q}|+\lambda)^2}-
\frac{\sin\theta_{\mathbf{q}}e^{\mp i\phi_{\mathbf{q}}}}{2ME-(\mathbf{K-k})^2-(\mathbf{K-p})^2-\mu(|\mathbf{q}|-\lambda)^2},\nonumber\\\label{Ab2}
\end{eqnarray}
where $\mathbf{q=k+p-K}$.

Note that we have assumed $\mathbf{K}$ is along $z$ axis. This means $|\mathbf{K-k}|$ and $|\mathbf{K-p}|$ are independent of $\phi_\mathbf{k}$ and $\phi_\mathbf{p}$. Moreover, it is easy to check that $|\mathbf{p}|$, $|\mathbf{q}|$ and $\theta_\mathbf{q}$ are only functions of $(\phi_\mathbf{k}-\phi_\mathbf{p})$, and $e^{ i\phi_{\mathbf{q}}}$ can also be expressed as a function in form of $F(\phi_\mathbf{k}-\phi_\mathbf{p})e^{ -i\phi_{\mathbf{k}}}$. Therefore, we find $Z_\updownarrow,S_\updownarrow,P_\updownarrow$ and $Q_\updownarrow$ should take following forms,
\begin{eqnarray}
Z_\updownarrow(\mathbf{K,k})&=&\tilde{Z}_\updownarrow(K,k,\theta_\mathbf{k}),\nonumber\\
S_\updownarrow(\mathbf{K,k})&=&\tilde{S}(K,k,\theta_\mathbf{k})e^{\pm i\phi_\mathbf{k}},\nonumber\\
P_\updownarrow(\mathbf{K,k,p})&=&\sum_{m}\tilde{P}_{\updownarrow m}(K,k,p,\theta_\mathbf{k},\theta_\mathbf{p})e^{im(\phi_\mathbf{k}-\phi_\mathbf{p})},\nonumber\\
Q_\updownarrow(\mathbf{K,k,p})&=&\sum_{m}\tilde{Q}_m(K,k,p,\theta_\mathbf{k},\theta_\mathbf{p})e^{im(\phi_\mathbf{k}-\phi_\mathbf{p})}e^{\pm i\phi_\mathbf{k}}.
\end{eqnarray}

And Eq.(\ref{Ab1}) become
\begin{eqnarray}
\left(
  \begin{array}{cc}
    \tilde{Z}_\uparrow & \tilde{S}e^{-i\phi_\mathbf{k}} \\
    \tilde{S}e^{i\phi_\mathbf{k}} & \tilde{Z}_\downarrow \\
  \end{array}
\right)
\left(
  \begin{array}{c}
    f_\uparrow \\
    f_\downarrow \\
  \end{array}
\right)=
\sum_m e^{im\phi_\mathbf{k}}
\int\frac{d^3\mathbf{p}}{(2\pi)^3}e^{-im\phi_\mathbf{p}}
\left(
  \begin{array}{cc}
    \tilde{P}_{\uparrow m} & \tilde{Q}_me^{-i\phi_\mathbf{p}} \\
    \tilde{Q}_me^{i\phi_\mathbf{p}} & \tilde{P}_{\downarrow m} \\
  \end{array}
\right)
\left(
  \begin{array}{c}
    f_\uparrow \\
    f_\downarrow \\
  \end{array}
\right).\label{Ab3}
\end{eqnarray}

If we decompose $f$ into $f(\mathbf{k})=\sum f^m(k,\theta_\mathbf{k})e^{im\phi_\mathbf{k}}$. It is clear that $f_\uparrow^m$ only couples to $f_\downarrow^{m+1}$. Therefore, we can write our ansatz as
\begin{eqnarray}
f_\uparrow(\mathbf{k})&=&f^m_\uparrow(k,\theta_\mathbf{k})e^{im\phi_\mathbf{k}},\nonumber\\
f_\downarrow(\mathbf{k})&=&f^{m+1}_\downarrow(k,\theta_\mathbf{k})e^{i(m+1)\phi_\mathbf{k}}.\label{Ab4}
\end{eqnarray}
Substituting the ansatz into Eq.(\ref{Ab3}), we obtain a two-dimensional integral equation,
\begin{eqnarray}
\left(
  \begin{array}{cc}
    \tilde{Z}_\uparrow & \tilde{S} \\
    \tilde{S} & \tilde{Z}_\downarrow \\
  \end{array}
\right)
\left(
  \begin{array}{c}
    f^m_\uparrow \\
    f^{m+1}_\downarrow \\
  \end{array}
  \right)
  =
  \int\frac{p^2\sin\theta_\mathbf{p}dpd\theta_\mathbf{p}}{(2\pi)^2}
\left(
  \begin{array}{cc}
    \tilde{P}_{\uparrow m} & \tilde{Q}_m \\
    \tilde{Q}_{m+1} & \tilde{P}_{\downarrow m+1} \\
  \end{array}
\right)
\left(
  \begin{array}{c}
    f^m_\uparrow \\
    f^{m+1}_\downarrow \\
  \end{array}
\right).\label{Ab5}
\end{eqnarray}

Although the above derivation of the ansatz(\ref{Ab4}) does not involve any symmetry argument. But it is clear that the decomposition of $f_\uparrow^m$ and $f_\downarrow^{m+1}$ is related to the second Casimir operator $\mathbf{P\cdot J}$ of SE(3) group. Since we have already assumed $\mathbf{K}$ is along $z$ axis, the conservation of $\mathbf{P\cdot J}$ is equivalent to the conservation of $J_z=L_z+s_z$. If we consider a bound state $|\Psi\rangle$ in the eigenspace of $J_z$, which means $J_z|\Psi\rangle=(m+\frac12)|\Psi\rangle$, it can be proved by a straightforward calculation that
\begin{eqnarray}
\Psi_\uparrow(\mathbf{k_1,k_2,k_3})&=&\delta(\mathbf{k_1+k_2+k_3-K})\sum_{m_1+m_2=m}\varphi_{\uparrow}^{m_1m_2}(k_1,\theta_1,k_2,\theta_2)e^{im_1\phi_1}e^{im_2\phi_2},\nonumber\\
\Psi_\downarrow(\mathbf{k_1,k_2,k_3})&=&\delta(\mathbf{k_1+k_2+k_3-K})\sum_{m_1+m_2=m+1}\varphi_{\downarrow}^{m_1m_2}(k_1,\theta_1,k_2,\theta_2)e^{im_1\phi_1}e^{im_2\phi_2}
\end{eqnarray}
Here $\varphi_\updownarrow^{m_1m_2}$ are undetermined functions that only depend on the magnitudes and polar angles of $\mathbf{k_1}$ and $\mathbf{k_2}$.

Recall that $f$ is defined by $f_\sigma(\mathbf{p})=g\sum_{\mathbf{q}}\Psi_\sigma(\mathbf{q,K-p,p-q})$. We can see that only $m_1=0$ terms contributes to $f_\sigma$, and we reproduce our ansatz after the summation over $\mathbf{q}$.

\section{Appendix C}

In this appendix, we will simplify Eq.(\ref{three_body_eqn_simplified}) with total momentum $\mathbf{K=0}$ and angular momentum $(J,J_z)=(j+\frac12,m+\frac12)$.

First, we write down Eq.(\ref{three_body_eqn_simplified}) more explicitly,
\begin{eqnarray}
Z_\uparrow(\mathbf{k})f_\uparrow(\mathbf{k})&+&S_\downarrow(\mathbf{k})f_\downarrow(\mathbf{k})=\\ \nonumber
&&\int\frac{d^3\mathbf{p}}{(2\pi)^3}
\bigg{[}\frac{1+\cos\theta_{\mathbf{k+p}}}{2M(E-\epsilon_\mathbf{k}-\epsilon_\mathbf{p}-\epsilon_\mathbf{k+p}^+)}+
\frac{1-\cos\theta_{\mathbf{k+p}}}{2M(E-\epsilon_\mathbf{k}-\epsilon_\mathbf{p}-\epsilon_\mathbf{k+p}^-)}
\bigg{]}f_\uparrow(\mathbf{p})\\ \nonumber
&&+\int\frac{d^3\mathbf{p}}{(2\pi)^3}
\sin\theta_{\mathbf{k+p}}e^{-i\phi_{\mathbf{k+p}}}
\bigg{[}\frac{1}{2M(E-\epsilon_\mathbf{k}-\epsilon_\mathbf{p}-\epsilon_\mathbf{k+p}^+)}+
\frac{1}{2M(E-\epsilon_\mathbf{k}-\epsilon_\mathbf{p}-\epsilon_\mathbf{k+p}^-)}
\bigg{]}f_\downarrow(\mathbf{p}),\\
Z_\downarrow(\mathbf{k})f_\downarrow(\mathbf{k})&+&S_\uparrow(\mathbf{k})f_\uparrow(\mathbf{k})=\\ \nonumber
&&\int\frac{d^3\mathbf{p}}{(2\pi)^3}
\bigg{[}\frac{1-\cos\theta_{\mathbf{k+p}}}{2M(E-\epsilon_\mathbf{k}-\epsilon_\mathbf{p}-\epsilon_\mathbf{k+p}^+)}+
\frac{1+\cos\theta_{\mathbf{k+p}}}{2M(E-\epsilon_\mathbf{k}-\epsilon_\mathbf{p}-\epsilon_\mathbf{k+p}^-)}
\bigg{]}f_\downarrow(\mathbf{p})\\ \nonumber
&&+\int\frac{d^3\mathbf{p}}{(2\pi)^3}
\sin\theta_{\mathbf{k+p}}e^{i\phi_{\mathbf{k+p}}}
\bigg{[}\frac{1}{2M(E-\epsilon_\mathbf{k}-\epsilon_\mathbf{p}-\epsilon_\mathbf{k+p}^+)}+
\frac{1}{2M(E-\epsilon_\mathbf{k}-\epsilon_\mathbf{p}-\epsilon_\mathbf{k+p}^-)}
\bigg{]}f_\uparrow(\mathbf{p}).
\end{eqnarray}
$Z_\updownarrow(\mathbf{k})$ and $S_\updownarrow(\mathbf{k})$ represent $Z_\updownarrow(\mathbf{K=0,k})$ and $S_\updownarrow(\mathbf{K=0,k})$ which have been defined in Appendix B.

After some straitforward calculation, we can simplify $S$ and $Z$ into

\begin{eqnarray}
Z_\updownarrow(\mathbf{k})&=&D(k)\pm W(k)\cos\theta_{\mathbf{k}} \\ \nonumber
S_\updownarrow(\mathbf{k})&=&W(k)\sin\theta_{\mathbf{k}}e^{\pm i\phi_{\mathbf{k}}},
\end{eqnarray}
with
\begin{eqnarray}
D(k)&=&-\frac{1}{2\pi(1+\mu)}\bigg{[}\frac{1}{a}\\ \nonumber&&-\frac{(k+\mu\lambda)\sqrt{\mu(k^2-2k\lambda-\mu\lambda^2)+(1+\mu)(k^2-2ME)}+
(k-\mu\lambda)\sqrt{\mu(k^2+2k\lambda-\mu\lambda^2)+(1+\mu)(k^2-2ME)}}{2(1+\mu)k}\bigg{]} \\
W(k)&=&-\frac{1}{12\pi(1+\mu)^2k^2}
\bigg{\{}\big{[}\mu(2k^2+k\lambda-\mu\lambda^2)+4k^2-2(1+\mu)ME\big{]}\sqrt{\mu(k^2-2k\lambda-\mu\lambda^2)+(1+\mu)(k^2-2ME)}\\ \nonumber
&&-\big{[}\mu(2k^2-k\lambda-\mu\lambda^2)+4k^2-2(1+\mu)ME\big{]}\sqrt{\mu(k^2+2k\lambda-\mu\lambda^2)+(1+\mu)(k^2-2ME)}
\bigg{\}}.
\end{eqnarray}
Then we can transform the integral equations into:

\begin{eqnarray}
\bigg{[}D(k)+W(k)\cos\theta_\mathbf{k}\bigg{]}f_\uparrow(\mathbf{k})+
W(k)\sin\theta_{\mathbf{k}}e^{-i\phi_\mathbf{k}}f_\downarrow(\mathbf{k})&=&
\int\frac{d^3\mathbf{p}}{(2\pi)^3}\bigg{[}M(\mathbf{k,p})+
N(\mathbf{k,p})(k\cos\theta_\mathbf{k}+p\cos\theta_\mathbf{p})\bigg{]}f_\uparrow(\mathbf{p})\nonumber\\
&+&\int\frac{d^3\mathbf{p}}{(2\pi)^3}N(\mathbf{k,p})(k\sin\theta_\mathbf{k}e^{- i\phi_\mathbf{k}}+p\sin\theta_\mathbf{p}e^{- i\phi_\mathbf{p}})f_\downarrow(\mathbf{p}),\nonumber
\end{eqnarray}
\begin{eqnarray}
\bigg{[}D(k)-W(k)\cos\theta_\mathbf{k}\bigg{]}f_\downarrow(\mathbf{k})+
W(k)\sin\theta_{\mathbf{k}}e^{i\phi_\mathbf{k}}f_\uparrow(\mathbf{k})&=&
\int\frac{d^3\mathbf{p}}{(2\pi)^3}\bigg{[}M(\mathbf{k,p})-
N(\mathbf{k,p})(k\cos\theta_\mathbf{k}+p\cos\theta_\mathbf{p})\bigg{]}f_\downarrow(\mathbf{p}) \nonumber\\
&+&\int\frac{d^3\mathbf{p}}{(2\pi)^3}N(\mathbf{k,p})(k\sin\theta_\mathbf{k}e^{ i\phi_\mathbf{k}}+p\sin\theta_\mathbf{p}e^{ i\phi_\mathbf{p}})f_\uparrow(\mathbf{p}),
\end{eqnarray}
with
\begin{eqnarray}
M(\mathbf{k,p})&=&\frac{1}{2ME+\mu\lambda^2-k^2-p^2-\mu(|\mathbf{k+p}|+\lambda)^2}
+\frac{1}{2ME+\mu\lambda^2-k^2-p^2-\mu(|\mathbf{k+p}|-\lambda)^2},\\
N(\mathbf{k,p})&=&\frac{1}{|\mathbf{k+p}|}
\bigg{[}\frac{1}{2ME+\mu\lambda^2-k^2-p^2-\mu(|\mathbf{k+p}|+\lambda)^2}
-\frac{1}{2ME+\mu\lambda^2-k^2-p^2-\mu(|\mathbf{k+p}|-\lambda)^2}\bigg{]}.
\end{eqnarray}

Noticing that $M$ and $N$ are functions of $k,p$ and $\cos\theta_\mathbf{kp}$, where $\theta_\mathbf{kp}$
is the angle between vectors $\mathbf{k}$ and $\mathbf{p}$. For $E<E_{\text{th}}$ and $\cos\theta\in[-1,1]$,
$M(\cos\theta),N(\cos\theta)$ are analytical. Therefore, we assume $M,N$ can be expressed as Taylor series of $\cos\theta$ like
$\sum a_n\cos^n\theta$. A Taylor series is a polynomial of $\cos\theta$, which means it can be written as
a series of Legendre polynomials like $\sum \tilde{a}_lP_l(\cos\theta)$. Using following addition theorem,
\begin{eqnarray}
P_l(\cos\theta_\mathbf{kp})=\frac{4\pi}{2l+1}\sum_{m=-l}^{m=l}Y_{l}^m(\Omega_{\mathbf{k}})Y_{l}^{-m}(\Omega_{\mathbf{p}}),
\end{eqnarray}
we conclude that $M$ and $N$ can be written in following forms:
\begin{eqnarray}
M(\mathbf{k,p})&=&\sum_{l,m}K_l(k,p)Y_{l}^m(\Omega_{\mathbf{k}})Y_{l}^{-m}(\Omega_{\mathbf{p}}),\\
N(\mathbf{k,p})&=&\sum_{l,m}R_l(k,p)Y_{l}^m(\Omega_{\mathbf{k}})Y_{l}^{-m}(\Omega_{\mathbf{p}}).
\end{eqnarray}

Finally, we obtain
\begin{eqnarray}
(D_\uparrow&+&W\cos\theta)f_\uparrow+W\sin\theta e^{-i\phi}f_\downarrow=\\ \nonumber
&&\int\frac{p^2dpd\Omega_\mathbf{p}}{(2\pi)^3}\sum_{l,m}Y_{l}^m(\Omega_{\mathbf{k}})Y_{l}^{-m}(\Omega_{\mathbf{p}})
\bigg{\{}[K_l+R_l(k\cos\theta_\mathbf{k}+p\cos\theta_\mathbf{p})]f_\uparrow+
R_l(k\sin\theta_\mathbf{k}e^{-i\phi_\mathbf{k}}+p\sin\theta_\mathbf{p}e^{-i\phi_\mathbf{p}})f_\downarrow\bigg{\}},\nonumber\\
(D_\downarrow&-&W\cos\theta)f_\downarrow+W\sin\theta e^{i\phi}f_\uparrow=\\ \nonumber
&&\int\frac{p^2dpd\Omega_\mathbf{p}}{(2\pi)^3}\sum_{l,m}Y_{l}^m(\Omega_{\mathbf{k}})Y_{l}^{-m}(\Omega_{\mathbf{p}})
\bigg{\{}[K_l-R_l(k\cos\theta_\mathbf{k}+p\cos\theta_\mathbf{p})]f_\downarrow+
R_l(k\sin\theta_\mathbf{k}e^{i\phi_\mathbf{k}}+p\sin\theta_\mathbf{p}e^{i\phi_\mathbf{p}})f_\uparrow\bigg{\}}.\label{3_body_eqn_2}
\end{eqnarray}

If we restrict the calculation in total translational momentum $\mathbf{K}=0$ and total angular momentum $(J,m_J)=(j+1/2,m+1/2)$
sub Hilbert space. The wave function in momentum space should take the form,
\begin{eqnarray}
\Psi_\uparrow(\mathbf{k_1,k_2,k_3})&=&\delta(\mathbf{k_1+k_2+k_3})\sum_{j_1,j_2,J'}\varphi_{j_1,j_2,J'}(k_1,k_2)\times \\ \nonumber
&&\sum_{m_1,m_2,m_J'}
\langle j_1,m_1;\frac{1}{2},\frac{1}{2}|J',m_J'\rangle\langle j_2,m_2;J',m_J'|j+\frac{1}{2},m+\frac{1}{2}\rangle Y_{j_1}^{m_1}(\Omega_\mathbf{k_1})
Y_{j_2}^{m_2}(\Omega_\mathbf{k_2}), \\ \nonumber
\Psi_\downarrow(\mathbf{k_1,k_2,k_3})&=&\delta(\mathbf{k_1+k_2+k_3})\sum_{j_1,j_2,J'}\varphi_{j_1,j_2,J'}(k_1,k_2)\times \\ \nonumber
&&\sum_{m_1,m_2,m_J'}
\langle j_1,m_1;\frac{1}{2},-\frac{1}{2}|J',m_J'\rangle\langle j_2,m_2;J',m_J'|j+\frac{1}{2},m+\frac{1}{2}\rangle Y_{j_1}^{m_1}(\Omega_\mathbf{k_1})
Y_{j_2}^{m_2}(\Omega_\mathbf{k_2}),\\ \nonumber
\end{eqnarray}
where $\langle j_1,m_1;j_2,m_2|J,M\rangle$ are Clebsch-Gordan coefficients.

Therefore, after summing over $\mathbf{k_1}$, only $j_1=m_1=0$ terms contribute to $f_\sigma$, and we obtain
our ansatz for $f_\sigma$,
\begin{eqnarray}
f_\uparrow(\mathbf{k})&=&\langle j,m;\frac{1}{2},\frac{1}{2}|j+\frac{1}{2},m+\frac{1}{2}\rangle f_0(k)Y_j^m
+\langle j+1,m;\frac{1}{2},\frac{1}{2}|j+\frac{1}{2},m+\frac{1}{2}\rangle f_1(k)Y_{j+1}^{m} \\ \nonumber
&=&\sqrt{\frac{j+m+1}{2j+1}}f_0Y_j^m-\sqrt{\frac{j-m+1}{2j+3}}f_1Y_{j+1}^m,\\\nonumber
f_\downarrow(\mathbf{k})&=&\langle j,m+1;\frac{1}{2},-\frac{1}{2}|j+\frac{1}{2},m+\frac{1}{2}\rangle f_0(k)Y_j^{m+1}
+\langle j+1,m+1;\frac{1}{2},-\frac{1}{2}|j+\frac{1}{2},m+\frac{1}{2}\rangle f_1(k)Y_{j+1}^{m+1}\\ \nonumber
&=&\sqrt{\frac{j-m}{2j+1}}f_0Y_j^{m+1}+\sqrt{\frac{j+m+2}{2j+3}}f_1Y_{j+1}^{m+1}.\nonumber\label{ansatz}
\end{eqnarray}
where $f_0$ and $f_1$ are two functions only depend on the magnitude of $\mathbf{k}$ and $Y_j^m$ is short for $Y_j^m(\Omega_\mathbf{k})$.

If we substitute this ansatz into Eq.(\ref{3_body_eqn_2}), we can get Eq.(\ref{final integral}) mentioned previously. And the coefficient matrices $Z$ and $K_j$ are given by
\begin{eqnarray}
Z(k)=\left(
       \begin{array}{cc}
         D(k) & -W(k) \\
         -W(k) & D(k) \\
       \end{array}
     \right),\qquad\qquad K_j(k,p)=\frac{p^2}{2\pi^2}\left(
                                     \begin{array}{cc}
                                       W_j(k,p) & -kR_{j+1}(k,p)-pR_j(k,p) \\
                                       -kR_j(k,p)-pR_{j+1}(k,p) & W_{j+1}(k,p) \\
                                     \end{array}
                                   \right).
\end{eqnarray}
\vspace{0.01in}
\end{widetext}

\end{document}